\documentclass[AMA,STIX1COL]{WileyNJD-v2}

\articletype{Survey paper}%

\received{15 March 2023}
\revised{6 June 2023}
\accepted{6 June 2023}

\usepackage{amsmath,amsfonts}
\usepackage{array}
\usepackage[caption=false,font=normalsize,labelfont=sf,textfont=sf]{subfig}
\usepackage{verbatim}
\usepackage{graphicx}
\usepackage{booktabs}
\usepackage{multirow}
\usepackage{tabularx}
\usepackage{xcolor}
\begin{document}
% Om Srirama Jai Vinayaka
\title{A Decade of Research in Fog computing: Relevance, Challenges, and Future Directions}

\author[1]{Satish Narayana Srirama $^{1}$}

\authormark{S. N. Srirama }
%\authormark{S. N. Srirama \textsc{et al}}

\address[1]{\orgdiv{School of Computer and Information Sciences}, \orgname{University of Hyderabad}, \orgaddress{\state{ Hyderabad 500 046}, \country{India}}}

% \address[1]{\orgdiv{School of Computer and Information Sciences}, \orgname{University of Hyderabad,}, \orgaddress{\state{ Hyderabad 500 046}, \country{India}}}

% \address[1]{\orgdiv{School of Computer and Information Sciences}, \orgname{University of Hyderabad,}, \orgaddress{\state{ Hyderabad 500 046}, \country{India}}}

% \address[1]{\orgdiv{School of Computer and Information Sciences}, \orgname{University of Hyderabad,}, \orgaddress{\state{ Hyderabad 500 046}, \country{India}}}

% \address[2]{\orgdiv{Org Division}, \orgname{Org Name}, \orgaddress{\state{State name}, \country{Country name}}}

% \address[3]{\orgdiv{Org Division}, \orgname{Org Name}, \orgaddress{\state{State name}, \country{Country name}}}

% \address[4]{\orgdiv{Org Division}, \orgname{Org Name}, \orgaddress{\state{State name}, \country{Country name}}}

\corres{Satish Narayana Srirama, \\ \email{satish.srirama@uohyd.ac.in}}

% \presentaddress{This is sample for present address text this is sample for present address text}

\abstract[Abstract]{

Recent developments in the Internet of Things (IoT) and real-time applications, have led to the unprecedented growth in the connected devices and their generated data. Traditionally, this sensor data is transferred and processed at the cloud, and the control signals are sent back to the relevant actuators, as part of the IoT applications. This cloud-centric IoT model, resulted in increased latencies and network load, and compromised privacy. To address these problems, \textit{Fog Computing} was coined by Cisco in 2012, a decade ago, which utilizes proximal computational resources for processing the sensor data. Ever since its proposal, fog computing has attracted significant attention and the research fraternity focused at addressing different challenges such as fog frameworks, simulators, resource management, placement strategies, quality of service aspects, fog economics etc. However, after a decade of research, we still do not see large-scale deployments of public/private fog networks, which can be utilized in realizing interesting IoT applications. In the literature, we only see pilot case studies and small-scale testbeds, and utilization of simulators for demonstrating scale of the specified models addressing the respective technical challenges. There are several reasons for this, and most importantly, fog computing did not present a clear business case for the companies and participating individuals yet. This paper summarizes the technical, non-functional and economic challenges, which have been posing hurdles in adopting fog computing, by consolidating them across different clusters. The paper also summarizes the relevant academic and industrial contributions in addressing these challenges and provides future research directions in realizing real-time fog computing applications, also considering the emerging trends such as federated learning and quantum computing.}

%This paper summarizes challenges, state-of-the-art and future research directions in realizing real-time fog computing applications. Contrary to other survey papers, that exhaustively address a specific set of aspects of fog computing, this work discusses the fog research challenges and solutions in much broader scope and thus provides a thorough opinion about progressing the research and quickly adapting fog computing in real-world applications.}


\jnlcitation{\cname{%
\author{S. N. Srirama} 
%\author{B. Hoskins}, 
%\author{R. Lee}, 
%\author{G. Masato}, and 
%\author{T. Woollings}
} (\cyear{2023}), 
\ctitle{A Decade of Research in Fog computing: Relevance, Challenges, and Future Directions}, 
\cjournal{Journal of Software: Practice and Experience}, \cvol{2023;00:1--20}.}

\maketitle

% \footnotetext{\textbf{Abbreviations:} ANA, anti-nuclear antibodies; APC, antigen-presenting cells; IRF, interferon regulatory factor}

\section{Introduction}\label{sec:intro}

Recent developments in the Internet of Things (IoT) and real-time applications, in domains such as smart cities, smart transportation, and smart healthcare etc., have led to the unprecedented growth in the connected devices such as embedded sensors, smart gadgets, smart phones, and industrial IoT tools etc. In the last two decades, the number of connected devices increased significantly, to over 50 billion, and the amount of data generated from them increased exponentially to the order of zettabytes. The amount of data generated from IoT connections worldwide is expected to increase further and reach 79.4 zettabytes by 2025~\cite{StatistaIoTData2022}. In the traditional model of Cloud-centric IoT~\cite{gubbi2013internet, chang2019internet},  this generated sensor data is moved to the cloud, where it is processed and the control signals are sent back to the relevant actuators, as part of the IoT applications. This migration of the data to the cloud for processing, increased the latency for these applications and lead to further challenges such as compromised privacy of the data and increased network load. To address these problems, \textit{Fog Computing} was coined by Cisco in 2012, a decade ago, which utilizes proximal computational resources for processing the sensor data~\cite{bonomi2012fog}.

Ever since its proposal, fog computing has attracted significant attention by both research community and industry, and people focused at addressing different challenges/aspects in realizing and utilizing the fog setup . These included fog architectures, geographical distribution, standardization efforts, resource provisioning, fog placement strategies, collaborative \& distributed data analytics, heterogeneity \& interoperability, device mobility, security \& privacy, scalability \& reliability, networking \& communication technologies and promoting real-time applications through fog economics. Lot of interesting results are obtained from these studies, which are applied in several prototypes for fog-based IoT applications. A search of Scopus database with “( ( fog  OR  edge )  AND  computing )” have resulted in over 48,000  publications.

However, after a decade of research, we still do not see large-scale deployments of public/private fog networks, which can be utilized in realizing interesting IoT applications. The market capture of fog computing is projected to be only around 343 million dollars by 2030~\cite{FogRevenue2030}, whereas, for comparison, the cloud market is projected to be around 791 billion dollars by 2028~\cite{CloudRevenue2030}. 

Furthermore, in the fog computing literature, we only see pilot case studies and small-scale testbeds, and utilization of simulators for demonstrating scale of the specified models addressing the respective technical challenges. There are several reasons for this lack of fog computing applications beyond small-scale pilot case studies. Most importantly, fog computing did not present a clear business case for the companies and participating individuals yet. Fog being perceived as cloud in proximity and strongly pushing opportunistic computing, the applications require significant infrastructure from nearby participants. However, there is not much literature about incentives to the participating individuals. Moreover, this type of offloading is perceived to be threat to the main business of proximal infrastructure providers such as mobile operators. This is one of the reasons for the Multi-access edge computing not being so successful yet. In addition, the energy requirements of the fog devices (sometimes battery powered) are significant and lacking models for efficiently utilizing the resources are further reasons for the sparsity of successful fog computing applications. 

\subsection{Scope and contributions}\label{sec:intro:scope}

This paper summarizes the technical, non-functional and economic challenges, which have been posing hurdles in adopting fog computing. The paper also summarizes the relevant academic and industrial contributions in addressing these challenges and provides future research directions in realizing real-time fog computing applications. The contributions of the paper are:

\begin{enumerate}

\item Fog computing challenges and the state-of-the-art (SOTA) are clearly organized into different clusters. % such as
\item Consolidated and presented the SOTA and ongoing research across these clusters. In the process also produced a consolidated figure (Figure~\ref{fig_3}) for fog computing challenges and solutions, which can be used as a quick reference in the domain.
\item Provides future research directions in realizing real-time fog computing applications also considering the emerging trends such as federated learning and quantum computing. 

\end{enumerate}

The rest of the paper is organized as follows:

We first explore the case studies of fog computing, justifying its relevance (Section~\ref{sec:rel}). We then explore the current research challenges, which are being addressed in fog computing domain, advancing the state-of-the-art. In the process, we also mention possible extensions to these studies, followed by a thorough discussion (Section~\ref{sec:challenges}). Later, we outline the future research directions in fog computing from the perspective of emerging trends in computer science, that could lead to the eventual deployment of fog networks and ubiquity of the fog-based applications (Section~\ref{sec:futureWork}). Section~\ref{sec:conclusion} provides a conclusion for the work.

%Contrary to other survey papers~\cite{yousefpour2019all, hu2017survey, laroui2021edge, gill2021comprehensive, bukhari2022fog, perera2017fog, goudarzi2022scheduling, mukherjee2018survey} in the fog computing domain, that exhaustively focus on a specific set of aspects of fog computing, this work discusses the fog research challenges and solutions in much broader scope and thus provides a thorough opinion about progressing the research and quickly adapting fog computing in real-world applications. 
%hu2017survey - removed this as it is generting one last ref
% @article{hu2017survey,
%   title={Survey on fog computing: architecture, key technologies, applications and open issues},
%   author={Hu, Pengfei and Dhelim, Sahraoui and Ning, Huansheng and Qiu, Tie},
%   journal={Journal of network and computer applications},
%   volume={98},
%   pages={27--42},
%   year={2017},
%   publisher={Elsevier}
% }

\section{Relevance of fog computing}\label{sec:rel}

\begin{figure}
\centering
\includegraphics[width=0.95\textwidth]{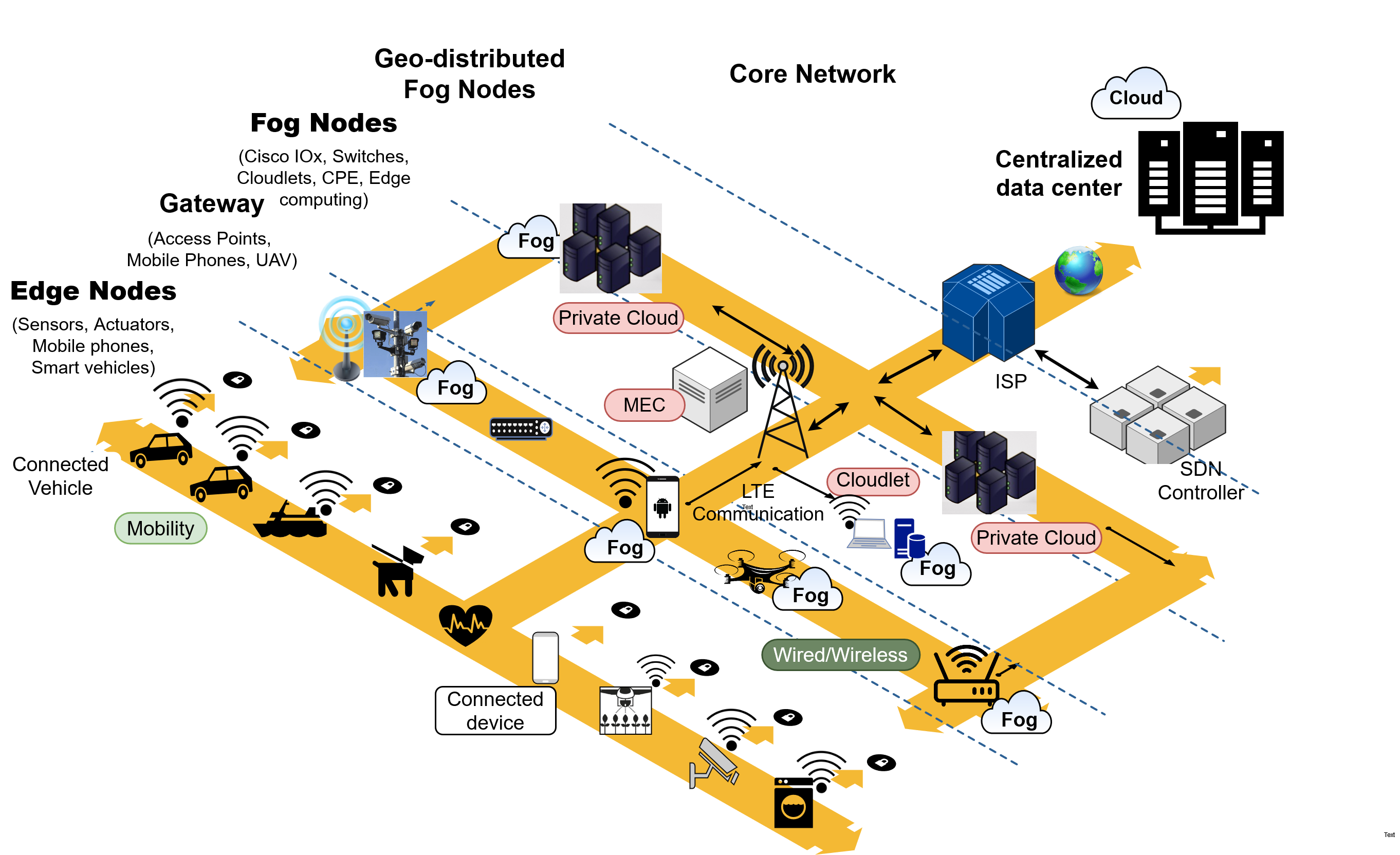}
\caption{Hierarchical fog computing architecture}
\label{fig_1}
\end{figure}

IoT primarily utilizes physical objects with sensing capabilities that are connected over the Internet, in realizing interesting smart applications. In these applications, instead of sending the raw sensor data from the edge devices to the cloud for processing, fog computing proposes utilizing computational resources available with the fog devices. The fog devices can be network switches, routers, gateway devices, desktops/laptops or even private clouds in proximity. Fog nodes can also be mobile and thus UAVs and smart phones can also be part of this fog network. Figure~\ref{fig_1} shows the hierarchical fog computing architecture with edge nodes, gateway devices, other fog nodes and centralized cloud. Processing anywhere from the edge nodes, until it reaches the centralized data centres, i.e. also on gateway devices, proximal and geo-distributed fog nodes, is considered to be part of fog computing. 

% \begin{figure}[t][width=0.49\textwidth]  //[width=0.49\textwidth,height=0.45\textwidth]
% \centerline{\includegraphics[width=342pt,height=9pc,draft]{empty}} [width=9cm,height=9cm]
% \caption{This is the sample figure caption.\label{fig1}}
% \end{figure}

This type of opportunistic computing by resource constrained edge devices is not a completely new idea. It was earlier studied in mobile web service provisioning (MWSP)~\cite{srirama2006mobile}, and mobile cloud computing that proposed cloudlets~\cite{satyanarayanan2009case} and code offloading~\cite{flores2015mobile}. MWSP offered services from resource constrained devices and smart phones, using standard web services communication technologies such as SOAP and REST (REpresentational State Transfer). This allowed them to capitalize on opportunistic computing in a peer to peer (P2P) manner. Cloudlets are micro data centres that provide cloud computing services to the smart phones. With code offloading, the mobile applications are profiled and offloaded to a much powerful surrogate in the cloud, under a particular context such as the current load, battery level, access to WiFi/mobile data etc. Similar concepts were also studied in Multi-access edge computing (MEC)~\cite{taleb2017multi} that utilizes the computational resources of mobile operators such as the ones with base stations. A detailed review of fog computing and its related computing paradigms, such as cloud computing, edge computing, mist computing etc. is provided at\cite{yousefpour2019all}.

\subsection{Fog computing case studies}

\begin{figure}
\centering
\includegraphics[width=0.85\textwidth]{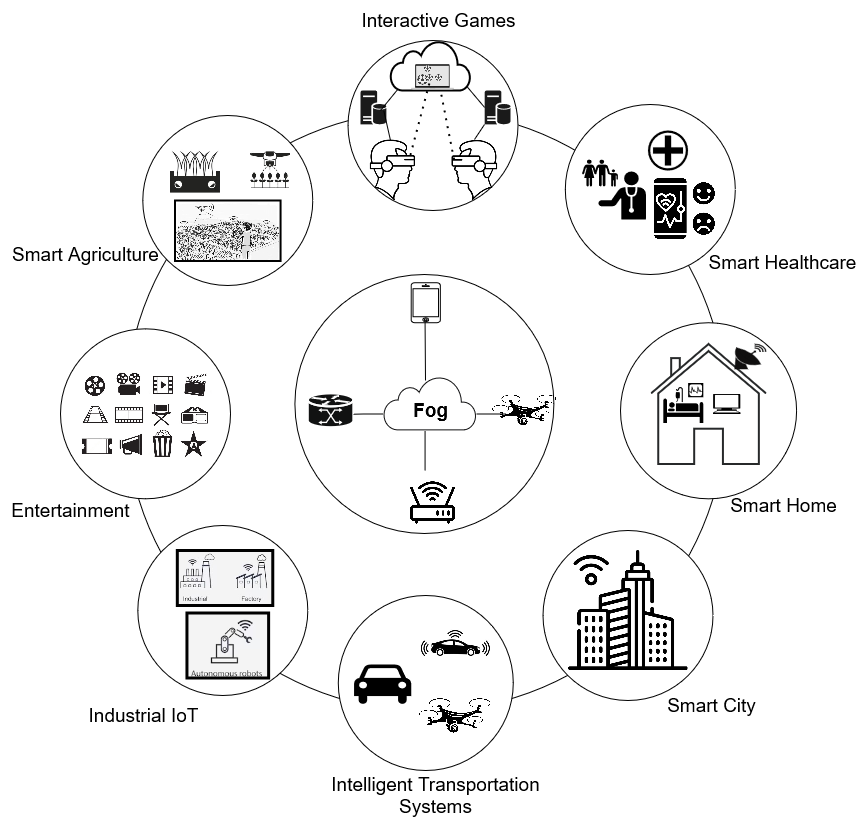}
\caption{Prominent fog computing application domains}
\label{fig_2}
\end{figure}

Fog computing is primarily shown to be relevant in scenarios that required real-time responses such as in smart homes, smart healthcare, interactive games, smart transportation, industrial IoT etc. Here are some of the prominent case studies/applications from the literature. Figure~\ref{fig_2} shows some of the prominent fog computing application domains. 

\subsubsection{Interactive games}

Some of the first demonstrated fog computing applications are related to interactive and virtual reality (IVR) games~\cite{zao2014pervasive, ha2014towards}. These applications are based on techniques such as augmented reality (AR) and brain monitoring, and required significant processing capabilities in the proximity and fog facilitated this without needing to send the collected gaming data to the centralized servers/clouds. With fog computing, only the consolidated results are exchanged among the participants for achieving real-time gaming experience.

\subsubsection{Smart healthcare}

This is one domain where lot of fog computing applications are demonstrated in the literature~\cite{kraemer2017fog}. Applications ranged from real time patient monitoring, elderly care to remote surgery. All these critical applications needed sub-second responses and fog computing with its proximal processing ability gave ideal options. Smart phones are also used here as fog nodes to collect and process the sensor data in proximity. Fog is also perfect in these scenarios as the medical data is confidential and, in most cases, it is preferred to store and process the data within the premises. 

\subsubsection{Smart home and smart city}

With the ubiquity of the smart gadgets, smart homes have improved the living standards of people significantly. Applications/case studies in this domain are mainly focusing on home automation, ambient assisted living, video surveillance and real-time monitoring. Fog computing is shown to facilitate several of these case studies through voice assistance, image/video and sensor data processing~\cite{rahimi2020fog}.  Similarly, in the domain of smart cities, fog computing case studies focused on smart water and waste management, real-time monitoring and control of city infrastructure such as traffic lights and power grids~\cite{perera2017fog}. Smart grids are modernized electrical grids that utilize IoT and AI (Artificial Intelligence) to improve the performance and efficiency of the grid, with fog computing facilitating the edge analytics. 

\subsubsection{Intelligent Transportation Systems (ITS)}

The original fog computing paper from CISCO in 2012~\cite{bonomi2012fog} already talked about fog facilitating the intelligent transportation through smart connected vehicles and smart traffic lights. These fog nodes equipped with sensors can detect the presence of pedestrians/cyclists and measure the speed of approaching vehicles. The data can be processed for accident prevention and maintenance of steady traffic. Later works based on fog computing focused on road maintenance, traffic management and city planning, as part of smart city applications~\cite{perera2017fog}. With the recent developments in autonomous vehicles and self-driving cars, fog computing can facilitate real-time video-aided navigation, vehicular communication, detection of hurdles and pedestrians, accident avoidance, etc. through computational, storage and network resources provided in proximity at roadside units~\cite{keshari2022survey}. 

\subsubsection{Industrial IoT}

The fourth industrial revolution, Industry 4.0, is focused on integrating advanced technologies such as IoT, AI, and cloud computing with industrial applications such as manufacturing and supply chain management. Fog applications are demonstrated, as part of IIoT, in different industries such as mining, smart grids, transportation, waste management, food processing etc., focusing on improving the productivity, operations and safety, by using different sensors, instruments and industrial equipment~\cite{basir2019fog}. Fog computing facilitated simultaneous data collection from variety of sensors, robots, and machines, pre-processing and interfacing incompatible sensors and machines through necessary protocol translation and mapping, as part of these applications~\cite{aazam2018deploying}.

\subsubsection{Entertainment}

Fog computing is also shown to be handy in improving the performance and scalability of entertainment applications, such as audio and video streaming. Fog nodes can be used to process and analyse these real-time video, multimedia and streaming data~\cite{satyanarayanan2015edge}. Fog facilitated finding best data streaming bit rate in different scenarios and adjusting video encoding rate (video processing speed) based on the current network load.

\subsubsection{Miscellaneous}
In addition to these case studies, fog computing is also demonstrated in other domains such as smart tourism, smart agriculture, environmental monitoring etc. In tourism applications, fog nodes were deployed across the city in assisting the visitors to find the next points of their interest. For environmental protection, unmanned aerial vehicles (UAV) and underwater robots were used as fog nodes, to collect and process the sensor data, in realizing scenarios such as coral reef monitoring and forest fire detection. In smart agriculture scenarios, drones as fog nodes are used to collect and process video and sensor data, in identifying pests and operating irrigation.

\subsection{Discussion}

From these case studies/applications one can deduce that fog computing is relevant and offers significant scope for smart applications in different domains. However, we do not see many real-world large-scale deployments of fog computing yet. We only encounter pilot case studies and small-scale testbeds based on fog computing. There are still several technical challenges that are to be fully addressed before fog computing can be realized seamlessly.

\section{Fog computing challenges and solutions} \label{sec:challenges}

\begin{figure}
\centering
\includegraphics[width=0.99\textwidth]{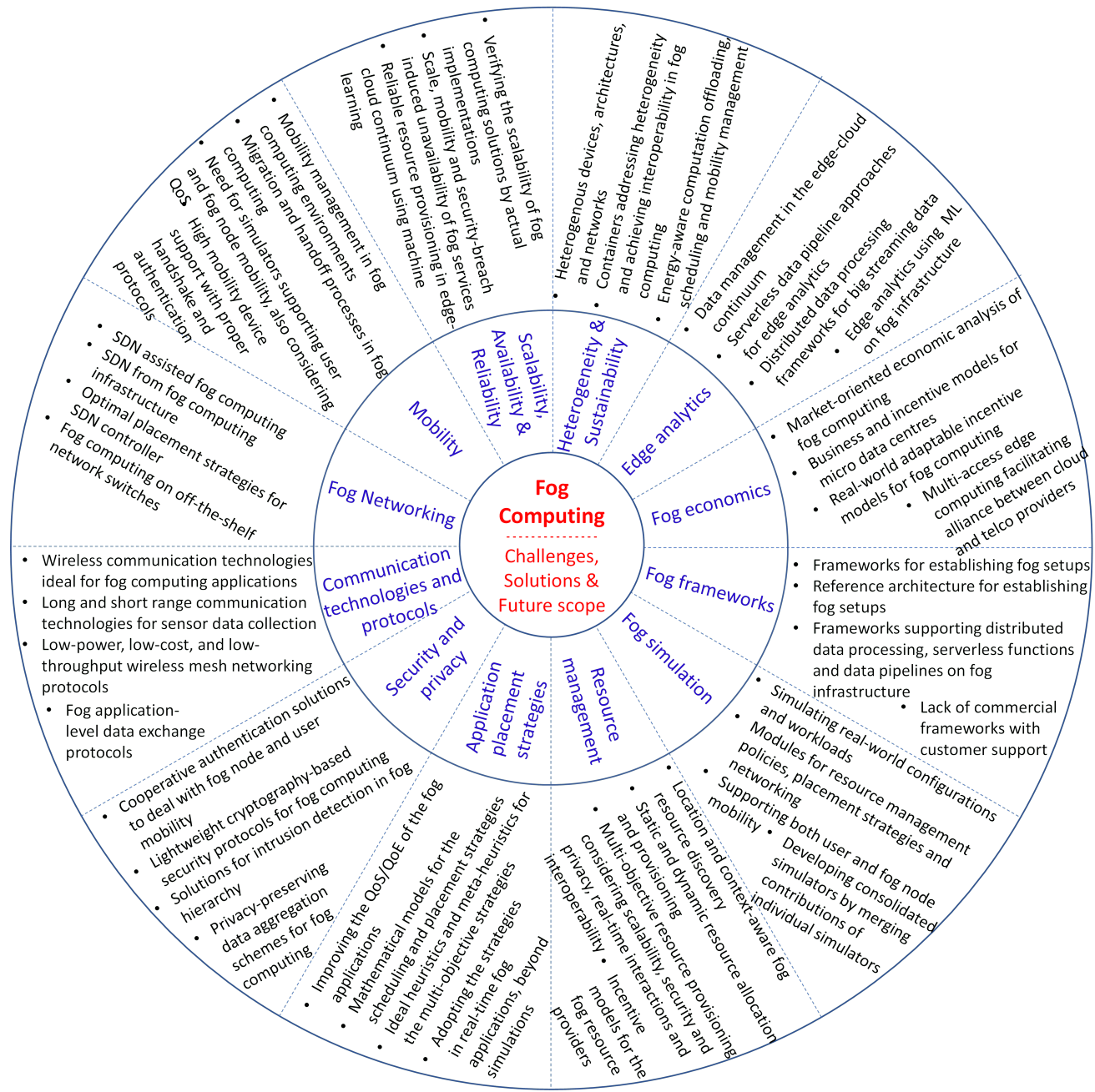}
\caption{Fog computing: challenges, solutions and future scope}
\label{fig_3}
\end{figure}

Fog computing poses several technical challenges, also including non-functional challenges, such as the need for ideal solutions for frameworks and simulators, resource management and scheduling, security and privacy, fog networking aspects, edge analytics, and issues with mobility, scalability, reliability, heterogeneity, and sustainability. There are also fog economics challenges that have been posing hurdles in adopting fog computing in real-world applications. Research in the IoT and fog computing domains tried to address these challenges, and here we try to summarize the respective challenges and identified solutions, by grouping them into different clusters. In the process we also discuss future extensions that are to be explored for these studies. Figure~\ref{fig_3} summarizes the fog computing challenges and solutions, by incorporating highlights of the discussion about the respective challenges, provided in the respective subsections (Sections~\ref{subsec:frameworks} - \ref{sec:cha:fogEconomics}). 

\subsection{Fog frameworks}
\label{subsec:frameworks}

With the initial interest generated in fog computing, people explored options for establishing fog setups. Several frameworks were proposed in the literature. Cisco proposed IOx as an application environment, on top of Cisco IOS and Linux OS kernel, that can be deployed on compatible hardware such as Cisco routers and switches, for providing fog services on resource constrained devices~\cite{CiscoIOx}. Later open source solutions such as Apache Edgent~\cite{ApacheEdgent} have appeared, proposing lightweight runtimes for streaming data processing on fog infrastructure. This interest and participation from multiple providers resulted in the formation of OpenFog Consortium, a public-private ecosystem, primarily targeted at accelerating the adoption of fog computing. OpenFog Consortium defined an open and interoperable reference architecture for establishing fog computing, which was later adopted by IEEE standards association~\cite{IEEEOpenFog}.

Several prototype fog frameworks were also proposed by different research groups that were used to demonstrate fog computing applications in different domains. Rahmani et al.  developed a prototype for fog-enabled health-care system~\cite{rahmani2018exploiting}. Chen et al.~\cite{chen2017smart} developed a fog framework for smart city surveillance. Generalized PaaS (Platform-as-a-service) models were also proposed for establishing fog setups such as Indie Fog, FogBus etc. Indie Fog~\cite{chang2017indie} proposed an approach/architecture for establishing fog setups using customer premise equipment such as idle desktops in cafes and handheld devices in proximity. FogBus~\cite{tuli2019fogbus} is a lightweight and distributed, container-based framework that can be used for integrating the IoT systems with edge, fog and cloud computing. Blockchain in FogBus ensures data integrity and the framework also supports other security aspects such as user authentication and data encryption. 

It is interesting to note that several of the fog frameworks are based on Docker containers, as the containers are light-weight enough to run on resource-constrained fog devices such as Raspberry PIs~\cite{morabito2017evaluating}. There are also frameworks which were targeted at executing specific type of fog applications. Srirama et al.~\cite{srirama2021akka}  proposed a framework that utilizes the proximal fog infrastructure for executing distributed computing applications using Actor programming model ~\cite{hewitt1973session} and containers. The framework is extended to CANTO~\cite{srirama2023canto}, that can be used to train neural networks on fog infrastructure for performing edge analytics. Frameworks were also developed to support executing serverless functions~\cite{golec2021ifaasbus} and data pipelines~\cite{poojara2022serverless} on fog infrastructure. The related works discussed in~\cite{srirama2023canto, tuli2019fogbus}, give a good summary of different fog frameworks. Most of these frameworks are research prototypes, mainly targeted at demonstrating novel applications in different domains. However, there are not many commercial frameworks with full customer support, for being able to adapt fog computing in real-time applications.

\subsection{Fog simulation}

While a lot of fog computing applications are demonstrated and research is shown in real devices, it is always required to study aspects such as scalability, scheduling, mobility etc. on simulators. The simulators allow us to study large scale fog deployments in much cost-efficient way. iFogSim~\cite{gupta2017ifogsim} was developed and extensively used for evaluating different resource management policies and placement strategies in fog computing, simulating real-world configurations and workloads. iFogSim extended the CloudSim, and thus inherited the core cloud computing features such as managing virtualization, applications, resources and scheduling functions. Later it was extended to iFogSim2~\cite{mahmud2022ifogsim2} with support for mobility, and microservice management in Edge and Fog computing environments. EdgeCloudSim~\cite{sonmez2018edgecloudsim}, is another tool based on CloudSim, that includes a nomadic mobility model, network model and an edge orchestrator for managing resources. MobFogSim~\cite{puliafito2020mobfogsim} is another simulator, which extended iFogSim and supports mobility and VM/container migration, to support the mobility of the consumer and thus live migrating the fog services to a different cloudlet. 

There are also simulators that focused on specific aspects of fog computing applications such as networking and business process management (BPM)~\cite{dumas2013fundamentals}. FogNetSim++~\cite{qayyum2018fognetsim++}, supports simulating different network characteristics of fog applications such as delay, packet loss, transmission range etc. and supports multiple communication protocols such as COAP, AMQP, MQTT, HTTP etc. along with support for different mobility models. Mobility and BPM are combined in STEP-ONE~\cite{mass2020step}, so that fog computing applications can be modelled as business processes. The processes can be executed on fog devices and can also be live migrated to other proximal fog nodes through offloading. A comprehensive review on simulators supporting fog computing is provided at~\cite{gill2021comprehensive}. 

While the existing fog simulators are interesting and sufficient for most of the general studies in fog computing, as stated already, each of the simulators is focused on specific aspects of fog computing such as support for scheduling, mobility, network, standards-compliant process management, etc. There are no efforts to merge all the individual contributions to come up with consolidated simulators that can be used in designing more complex fog computing scenarios. This is also making further contributions in the simulators more fragmented. 

\subsection{Resource management}

Fog computing applications can be of two types: 1. Top-down applications 2. Bottom-up applications. In top-down applications, cloud provider deploys the necessary geo-distributed fog resources for managing the execution of fog applications across the fog topology. The provisioning and management of resources can be based on different QoS (Quality of Service) parameters such as proximity, user load, latency, cost models etc. In the bottom-up approach, applications are managed by individual fog service providers. Here the fog resources are provided by different vendors. These proximal resources are identified and fog related tasks are scheduled, by the gateway devices. Thus, the onus of resource discovery, provisioning, scheduling and placement fall on the fog application/service provider. With both the types of fog applications, resource management is a huge challenge and significant literature~\cite{ghobaei2020resource, hong2019resource} tried to address this challenge during the past decade. 

Fog computing generally deals with storage, processing and network resources. First the relevant resources and fog nodes providing them are to be identified. The resource discovery solutions should be location-aware, to reduce the geographic and network distances. They also should be context aware considering the fog application requirements and current load and resource availability of proximal fog nodes. Bukhari et al.~\cite{bukhari2022fog}, provides a detailed literature review of fog node discovery and selection. Regarding the fog resource allocation and provisioning, the approaches can be static or dynamic. In the static provisioning, the resources of the fog nodes are fixed/known in advance, and the tasks of multiple fog applications are scheduled on the fog nodes based on different objectives such as energy-efficiency, reducing latency, bandwidth usage and cost. In the dynamic approaches, the fog providers tried to bring in additional resources based on the need. These studies involved extensive cost models, auction models etc. again considering the multi-objectives. Ghobaei-Arani et al.~\cite{ghobaei2020resource}, provides a comprehensive review of the approaches for resource allocation and provisioning. Regarding resource scheduling and application placement, the main goal is to find the best viable assignment of available resources based on fog application requirements, which is discussed in detail in the next subsection. A detailed survey of architectures and algorithms, considered for resource management are provided at~\cite{hong2019resource}.

While significant literature already tried to address this challenge, there is still a huge scope for future work. Mainly, the approaches should focus at multi-objectives also considering other QoS aspects such as scalability, privacy and security, real-time interactions and interoperability of the devices. Even though cost models are studied extensively, there is still not enough and ideal incentive models for the fog resource providers (Section~\ref{sec:cha:fogEconomics}). This is one of the main hurdles in achieving large scale fog computing deployments in the real-world. 

\subsection{Application placement strategies}

The main goal of the fog application scheduling and placement is to find the best feasible assignment of available resources to fog application requirements. Scheduling fog applications on the infrastructure and thus the fog placement strategies are extensively studied in the literature~\cite{goudarzi2022scheduling, salaht2020overview, brogi2020place}. The approaches mainly focused at improving the QoS/QoE of the fog applications. In terms of QoS, the strategies tried to reduce the latency, cost and energy utilization~\cite{goudarzi2022scheduling}. In addition to QoS parameters, the QoE (Quality of Experience) approaches considered further user perspective parameters such as required access rate, priority of applications and processing time~\cite{mahmud2019quality}. The fog applications are modelled as i) monolithic, i.e., single program for the application ii) Independent, i.e., set of independent tasks executing for the application iii) Modular, i.e., set of dependent tasks executing at different locations constituting the full fog application. Approaches were studied to place all the three models of fog applications on both homogeneous and heterogeneous infrastructure, considering the objectives from both the fog user (e.g. deadline and priority-aware~\cite{adhikari2019dpto}) and provider (e.g. efficient utilization of resources and maximizing the profit~\cite{bahreini2021mechanisms, mahmud2020profit}) perspectives. 

The fog placement strategies in the literature are based on single objective or multi-objective. The problems were mostly modelled as Integer Linear Programming (ILP) (e.g.~\cite{ma2021towards}), Mixed Integer Linear Programming (MILP) (e.g.~\cite{peng2021constrained}) and Markov Decision Process (MDP) (e.g.~\cite{goudarzi2021distributed}). Since these multi-objective strategies are NP-hard, several heuristics (e.g. constant-factor approximation~\cite{farhadi2021service}) and meta-heuristics (e.g. based on Genetic Algorithm (GA)~\cite{guerrero2019evaluation}, Simulated Annealing (SA)~\cite{fang2021sap}, Particle swarm optimization (PSO)~\cite{adhikari2019application}) were proposed. Nature-Inspired meta-heuristics solutions for placement and scheduling of fog applications are summarized in~\cite{adhikari2022comprehensive}. 

In addition to these approaches, Machine Learning (e.g. Reinforcement Learning (RL)~\cite{tao2021adaptive}, Residual Recurrent Neural Networks~\cite{tuli2020dynamic}) and game theory (e.g. Stackelberg Game~\cite{hazra2020stackelberg})  were also applied for efficiently scheduling the applications across the fog topology. Goudarzi et al.~\cite{goudarzi2022scheduling} have nicely summarised the fog placement strategies in detail. 

However, most of the fog placement solutions were proposed on simulators and very few of them  are actually demonstrated in small-scale practical implementations. There is still a significant scope to address the challenge in realizing placement strategies in real-time fog applications. 

\subsection{Security and privacy}
\label{sec:cha:security}

Since fog computing mainly provides computation, storage and network services on resource-constrained devices at the edge of the network, it faces several security and privacy breaches. Generally, fog is considered to be more secure than cloud computing. With fog, reliance on Internet decreases as the sensor data is accumulated and analysed locally. This eliminates threats such as eavesdropping and man-in-the-middle attacks over the internet. The privacy of the user data also gets improved as the data do not leave the premises. However, the communication between the edge and fog devices is more prone to attacks such as tampering, jamming and denial-of-service~\cite{ni2017securing}. The problem is alleviated further, since due to the resource-constrained nature of the fog devices, not many standard security protocols such as encryption and public-key cryptography can be applied for the communication. 

Several works focused at addressing security and privacy issues in the fog. Primarily, fog nodes assisted in authenticating the users in several IoT applications~\cite{hu2017security}. However, the decentralization of fog computing, and mobility of users and fog nodes makes this identity authentication a complex issue. Cooperative authentication solutions are proposed to address this challenge~\cite{lin2013achieving}. Regarding security protocols in fog computing, lightweight cryptography is proposed, whose properties are discussed in ISO/IEC 29192, with block cipher solutions such as PRESENT, CLEFIA and LEA~\cite{ISOLightweightCrypto, yoshida2014status}. However, there are no public-key cryptography approaches based on this lightweight cryptography, due to the inherent computational complexity of public-key cryptography. Therefore, regarding transient data storage on fog nodes and to enable confidential and privacy-preserving data sharing among the participating nodes, solutions were proposed based on efficient key exchange protocols~\cite{alotaibi2017attribute} and homomorphic encryption~\cite{lu2017lightweight}.

Another major security challenge in fog computing is the intrusion detection, as malicious internal and external attackers can hack different entities in fog hierarchy. Different anomaly-based, machine learning-based and statistical-based techniques for intrusion detection are proposed in the literature. Intrusion detection systems in the fog environment are thoroughly discussed in~\cite{yi2023deep}.  Solutions are also proposed to ensure end-to-end trustworthiness in fog computing based on security attributes of all the participating nodes~\cite{su2017toward}. A detailed review of security in fog computing is provided at~\cite{ni2017securing}. Solutions based on blockchains addressing security, privacy, distributed trust management, and reliability challenges in fog computing are summarized in ~\cite{alzoubi2022blockchain}. However, with security being a non-functional requirement and new threats appearing in Internet on regular basis, it is obvious that security in fog computing is a continuous research with significant future scope. 

\subsection{Communication technologies and protocols}

In the hierarchical fog computing architecture (Figure~\ref{fig_1}), with edge nodes, fog nodes and cloud data centre along the edge-cloud continuum, there is scope for three kinds of connections/communication among the devices. 1. Wireless connection between the edge devices and the fog nodes; 2. Wired/wireless connection among the fog nodes; 3. Wired/wireless connection between the fog nodes and the cloud data centre. There are several communication technologies and protocols, which evolved over the years, that actually make these connections/communication feasible. The technologies are specifically focused on optimizing energy-efficiency, communication range, standards compliance, bandwidth etc., thus facilitate the IoT/fog applications. Perera et al~\cite{perera2017fog}, provides a detailed summary of the communication technologies used in fog computing domain. The paper also provides information on which network layer of Open Systems Interconnection (OSI) model, each of the communication technologies belongs to. 

Mainly for wireless communication over the physical and data link layers, technologies such as WiFi, Bluetooth, Bluetooth Low Energy (BLE) are commonly used for communication among the edge and fog nodes. Near field communication (NFC) and Radio-frequency identification (RFID) are used for communication over short distances ranging from few centimetres to meters. For communication over the long distances (ranging 2-50 KM in urban and suburban regions), technologies such as Long Power Wide Area Network (LPWAN or LoRaWAN) or Sigfox can be used. These protocols are low-powered, low-cost, and low-bit rate, specifically designed for two-way secure communication in the IoT domain. Mobile communication technologies such as 4G and 5G are also used for this wireless communication, when the smart phones act as gateway devices and fog nodes. It is also interesting to note here that fog computing based radio access network~\cite{peng2016fog} is shown to provide high spectral and energy efficiency in 5G system.

For wireless communication over higher layers (network, transport and application layers) protocols such as ZigBee, Z-Wave are used. ZigBee is one of the most popular low power, cost, and throughput, wireless mesh networking standard. ZigBee needs an application-level gateway to connect to the Internet using Ethernet or WiFi. Z-Wave is designed to facilitate device to device (D2D) communication in smart home applications. Similarly, 6LoWPAN (IPv6 over Low-power Wireless Personal Area Networks) enables IP based communication in low-powered devices~\cite{perera2017fog}. 

Regarding wired communication, standard TCP/IP protocols are used over Ethernet. Specifically to support fog computing, adapting Long-Reach Passive Optical Network (LRPON) is proposed, which makes the network reach up to 100 km~\cite{zhang2017infrastructure}. Further developments regarding fog networking challenge are discussed in the next subsection. 

Regarding application-level data exchange in IoT/fog applications, protocols such as Hypertext Transfer Protocol (HTTP), Message Queue Telemetry Transport (MQTT), Extensible Messaging and Presence Protocol (XMPP), Constrained Application Protocol (CoAP), Advanced Message Queuing Protocol (AMQP), etc. are used. Donta et al~\cite{donta2022survey}, provides a detailed survey on IoT application layer protocols. While there exist several options, choosing ideal communication technologies and protocols for a specific fog application is a major challenge.

\subsection{Fog Networking}

The traditional networking architectures and protocols of the Internet were not designed for the high-level scalability demands of IoT. The billions of connected devices and the zettabytes of sensor data moved to the cloud for processing, result in congestion in the core network. Fog computing addresses this issue by processing the data in proximity. However, it is important to note that augmenting the computational capabilities of fog nodes is not a complete replacement for the cloud. Cloud and fog computing are complementary in nature and need to be employed together in different applications. However, it is not straightforward to employ cloud and fog together while letting them accessible ubiquitously. A minimum of three tier architecture is necessary along with its own set of coordination and orchestration requirements. Thus, an intermediate networking layer is required to orchestrate the communication between fog and cloud servers, the communication among the fog devices in a device-to-device (D2D) manner, and to achieve seamless service delivery and handover mechanisms to support mobility of fog nodes. Software defined networking (SDN) is an emerging computing and networking paradigm, that can be used for the orchestration of this intermediate networking layer~\cite{baktir2017can}. 

SDN separates control plane and data plane to realize the flexible control of network traffic. A centralized server takes care of the control, and thus network routing and transmission rules can be defined at this centralized node making communication more flexible and intelligent~\cite{kreutz2014software}. This allows network switches to utilize all their hardware resources for just forwarding data rather than also using them for computing routes. OpenFlow, an open protocol, provides a standard way of communication between controller and switch, allowing reprogramming and updates in the FlowTables. SDN assisted fog computing is studied extensively in the literature. Tomovic et al.~\cite{tomovic2017software} proposed an SDN-assisted control and monitoring framework for the fog-based IoT network. Baktir et al.~\cite{baktir2017can} provides a detailed study about how fog/edge computing can benefit from SDN. Primarily, with SDN controller, a global view of the fog network is available with information such as available resources (memory and storage of fog nodes) and software applications. This information can be used in delivering fine-grained QoS provisioning for fog services~\cite{huang2016software}. Xu et al.~\cite{xu2016towards} joined SDN and MQTT protocol for effective and reliable delivery of IoT data. The paper also showed that SDN can effectively operate within the fog computing infrastructures. Here the SDN controller is placed on a fog node that acts as a broker for MQTT clients. % The basic architecture of fog computing with SDN is discussed in [] . 

The SDN control and other functionality across the fog hierarchy can be deployed through Network Functions Virtualization (NFV)~\cite{mijumbi2015network}. By leveraging the virtualization technology with NFV, the network functions are decoupled from the dedicated physical network hardware. Thus, the gateways, switches, and firewalls can be virtualized and placed as fog nodes, where the SDN functionality can also be deployed. NFV technology is shown to improve the flexibility of telecommunication service provisioning, with solutions such as TelcoFog controller~\cite{vilalta2017telcofog}. SDN/NFV based fog/edge computing solutions are discussed further in~\cite{laroui2021edge}.

While SDN controller achieves efficient management of heterogeneous fog networks, the optimal placement of the coordinator is still an important design issue. Scalability issues of the controller are also to be considered and to address this, distributed SDN networks are proposed. In this approach physically independent SDN controllers manage a set of subnetworks, which synchronize among themselves to achieve a logically centralized network view. Hakiri et al.~\cite{hakiri2017managing}  proposed an SDN-enabled wireless fog architecture, with a hybrid SDN control plane. However, there is still a significant scope for extending the study. Furthermore, implementing fog computing in off-the-shelf network switches is an active research topic~\cite{mukherjee2018survey}.

\subsection{Mobility}

Mobility is a one of the most interesting challenges relevant for fog computing, as several of the fog devices, such as smart phones, vehicles, and drones are highly mobile. In fog applications, both the edge devices that produce the sensor data and act as the fog users, and the fog devices where the intermediate processing is being performed, can be mobile. For example, in applications such as autonomous vehicles and ITS, the vehicles are mobile and roadside units can act as static fog nodes. Similarly, in applications that require extracting sensor data from remote locations, a drone can act as a fog node, which can collect and process the data. In both the types of scenarios, mobility of edge or fog nodes impairs fog computing performance. 

Mobility mainly causes a change of the access points. Therefore, the data and current services being processed should migrate to the fog node at the new access point. Several works tried to address this mobility management in fog computing environments~\cite{zaman2021mobility, waqas2018mobility, pereira2019assessing, zhang2017fog}. For example, Ghosh et al.~\cite{ghosh2019mobi}, proposed a mobility aware collaborative framework, Mobi-IoST. Here, when the mobile user changes its location, a centralized node is informed, which analyses the user mobility patterns and with the help of a Markovian model, predicts the user’s next location. When the user reconnects with the new fog node, the fog services are migrated to this new node, and the results are finally delivered to the mobile user. Mobi-IoST, framework reveals the steps in fog mobility management, which are further categorized in~\cite{puliafito2020mobfogsim}. The intermediary steps in this mobility management procedure are explored extensively, also leaving scope for further work. For example, migration and handoff processes in fog computing are thoroughly examined in~\cite{rejiba2019survey}.

Mobility management in fog computing is generally explored through simulation, as it includes too many dynamic parameters, which cannot be fully observed through real-time applications. MobFogSim~\cite{puliafito2020mobfogsim} is a simulator that supports mobility and VM/container migration. Mobility and BPM are combined in STEP-ONE~\cite{mass2020step}, which supports modelling fog computing applications as business processes and the processes can be live migrated across fog/edge nodes. Gill and Singh~\cite{gill2021comprehensive}, in the process of providing a comprehensive review of fog simulators, categorized all the simulators that support mobility. 

Mobility significantly decreases the QoS (such as latency, scalability, reliability etc.) and the security of the fog applications. Thus, the high mobility device support in fog computing is a very important issue that needs to be explored further. Moreover, it is necessary to develop handshake and authentication protocols, which are quick and stateless, for supporting high-speed users and automotive communication~\cite{zhang2017cooperative}.

\subsection{Scalability, Availability \& Reliability}

Many of the existing algorithms and schemes addressing different challenges of fog computing, do not scale well to the magnitude of IoT networks, with billions of connected devices. The studies neglect scalability in their fundamental design or do not have access to enough resources, when they are tested in real devices. For example, very few of the pilot case studies discussed in Section~\ref{sec:rel} have considered scalability as an intrinsic part of the application. Scalability in literature is mostly studied through simulators, assuming certain environmental conditions. Several critical parameters may be missing in these simulations and only large scale deployments of fog networks and infrastructure can prove the adaptability of these solutions in the real-world. Thus, the research community is strongly encouraged to verify the scalability of their proposed solutions by actual implementations~\cite{yousefpour2019all}. 

The scalability of IoT/fog networks leads to the issues with the availability. Not enough resources may be available with the proximal fog nodes to support all the fog users and some of them may be left with no fog nodes to offload. This problem is addressed in the literature dealing with resource management challenge~\cite{ghobaei2020resource, hong2019resource}. Mobility also leads to unavailability issues. The edge device/fog user might have moved away, and the results cannot be delivered after processing is finished at the fog node, and in the other case, the fog node might have moved away after the tasks are delivered to it~\cite{zaman2021mobility}. Security also intensifies this availability issue, with attacks such as denial of service (DoS), and the studies relevant to security challenge (Section~\ref{sec:cha:security}) are addressing this issue. 

The corollary of scalability and availability challenges of fog computing is the issue with reliability. Reliability aspects of fog computing deal with both failures and recoveries. Reliability of fog computing is studied in the literature in different contexts. Yao and Ansari~\cite{yao2019fog}, studied fog resource provisioning in reliability-aware IoT networks. Hou et al.~\cite{hou2020distributed}, tried to address the reliability while collectively utilizing the fog nodes for distributed computing. They designed the task allocation as an optimization problem, considering the latency, reliability and energy consumption. Desikan et al.~\cite{desikan2022decoding}, studied the relationship among latency, reliability, cost, and energy, while provisioning resources in fog networks. Duc et al~\cite{duc2019machine}, provides a detailed review on reliable resource provisioning in edge-cloud continuum using machine learning, with significant scope for future work.

\subsection{Heterogeneity \& Sustainability}

IoT deals with heterogeneity everywhere, with heterogeneous devices, heterogeneous architectures, and heterogeneous networks~\cite{qiu2018can}. Heterogeneity of edge nodes are addressed in fog computing through solutions such as “IoT Hubs”, that act as a bridge between the different physical networks and merging them using an all-IP network~\cite{cirani2015iot}. Fog nodes are also heterogeneous, as they can be network switches, routers, gateway devices, and proximal computational resources. Thus, they may have different hardware architectures and features, software compatibility, supported communication protocols, energy requirements, and processing and storage capabilities. However, the fog users should be able to take advantage of fog resources through seamless offloading. Hong et al.~\cite{hong2013mobile}, proposed Mobile Fog, a platform as a service (PaaS) model that provides a programming abstraction and allows applications to use heterogeneous fog resources while also supporting dynamic scaling at runtime. Similar PaaS models were also proposed in solutions such as~\cite{yangui2016platform}. Later containers were explored to be executed on fog infrastructure to address the heterogeneity. Containers virtualize the operating system and run anywhere, including resource constrained edge/fog devices such as Raspberry PI, thus provide an ideal solution for offloading~\cite{kaur2017container}. Standardized container solutions such as Docker containers helped in achieving interoperability in fog computing. 

Heterogeneity of fog devices is also considered in addressing other challenges such as placement strategies. E.g. In Kattepur et al.~\cite{kattepur2017priori}, computational times of conventional robotic runtime algorithms are estimated on heterogeneous hardware, to decide whether to offload the tasks to proximal fog node or not. Research works addressing heterogeneity in fog computing are categorized in Appendix 1 of ~\cite{yousefpour2019all}.

Sustainability challenge primarily deals with energy consumption of the devices in the fog networks. Heterogeneous fog nodes and edge/IoT devices may be deployed in locations such as in forests, minefields, underwater etc., for realizing different smart scenarios. Thus, the nodes may be battery powered or depend on renewable energy sources (such as solar, wind, or vibration). Therefore, it is critical to conserve energy in these edge/fog devices, while addressing different challenges of fog computing. Energy-efficiency is considered in addressing fog computing challenges such as energy-aware computation offloading (e.g.~\cite{materwala2022energy}), energy-aware scheduling (e.g.~\cite{jayanetti2022deep}), energy-aware mobility management (e.g.~\cite{zhang2015optimal}) etc. Future work in this domain should focus on energy harvesters and battery storage, for sensors and edge devices. It is also interesting to identify where and how close the fog nodes should be placed to the end users, to optimize the energy usage.

\subsection{Edge analytics}

In the edge-cloud continuum, the zettabytes of sensor data collected from the things/devices is analysed, interpreted and presented, across the layers in realizing interesting IoT/fog applications. Data management in the edge-cloud continuum and edge analytics include data integration (propagation, federation and consolidation), storage, preprocessing (filtering, anonymization, error detection), batch and stream processing, and provisioning. Storing and delivering data across fog hierarchy is studied in detail and solutions such as Data as a Service have appeared~\cite{plebani2018fog, badidi2020fog}.

The processing of the data across the edge and fog devices is termed as edge analytics, which includes preprocessing and stream data processing~\cite{satyanarayanan2015edge}. The data can be perceived to be executed in a pipeline of processes, where each process takes care of certain task such as collection, filtering, rule-based propagation across several paths, merging etc. The processes can also be based on serverless computing, an event-driven way of invoking functions. Serverless data pipeline approaches for edge analytics are evaluated in~\cite{poojara2022serverless}. Edge analytics provides early insights from data streams and protect data stores at cloud against massive data volumes, high data velocity and network congestion. The fog infrastructure can also be used to perform distributed data analytics on the collected IoT data, by collectively utilizing the storage and processing capability of the fog nodes~\cite{srirama2021akka}. 

Distributed data analytics is better performed at the clouds, due to the availability of unlimited processing power~\cite{assunccao2015big}. Distributed data processing frameworks such as Hadoop MapReduce and in-memory alternatives such as Apache Spark, can be employed for this sensor data analytics on the cloud. Moreover, since IoT mostly deals with \textit{big streaming data}, message queues such as Apache Kafka can be used to buffer and feed the data into stream data processing systems such as Apache Storm and Apache Spark streaming~\cite{yang2017iot}.

Additionally, the edge analytics performed can be based on machine learning (ML). E.g. Drolia et al.~\cite{drolia2017precog}, proposed a PreCog system on fog devices that recognizes images rapidly through catching and prefetching. Abdulkareem et al.\cite{abdulkareem2019review}, provides a detailed review of approaches performing edge analytics using  ML on fog infrastructure. The ML models and their corresponding functions such as clustering, classification and feature extraction, in the context of IoT are extensively investigated in~\cite{mahdavinejad2018machine}. There are some generic distributable algorithms such as k-nearest neighbors (k-NN) and other special neural network methods, which can directly be used in resource constrained fog devices for performing ML tasks. These studies lead to the development of frameworks such as CANTO~\cite{srirama2023canto}, that can be used to train neural networks on fog nodes for performing edge analytics. However, performing sophisticated ML algorithms in resource and power constrained fog nodes is still a major challenge. Support for ML enabling hardware such as ENVISION, that can be used in fog networks, is summarized in~\cite{zou2019edge}.

\subsection{Fog economics}
\label{sec:cha:fogEconomics}

Fog computing evolved with the aim of bringing cloud services (storage, computing and networking) to the proximity of IoT/edge devices. This will result in reduction of network transport costs and latency, by processing the data where it is generated. Latency reduction can benefit user experience and thus can drive revenue growth, and indirectly improves labour productivity as more customers/applications can be supported~\cite{weinman2017economics}. Moreover, with the distributed fog, the risk of total loss of service is eliminated, as the complete dependence on centralized cloud is minimized.

Since fog is cloud in proximity, the \textit{economics of cloud computing} challenge and its relevant studies, to some extent, are directly applicable to fog computing. Cloud economics focused at 1. Pricing of cloud services 2. Brokerage mechanisms providing access to appropriate cloud resources/services based on user requirements 3. Monitoring to determine if proper SLAs (Service Level Agreements) are guaranteed~\cite{buyya2018manifesto}. However, fog economics is not explored considerably in the literature, in any of the three directions, leaving significant scope for future work. Very few works exist and Kim et al.~\cite{kim2019economics} performed theoretical economic analysis of fog computing with a market consisting of Infrastructure and Service Provider (ISP) as brokers, fog users, and Edge Resource Owners (EROs). The problem is designed as a non-cooperative game.

In addition, the original architectures of fog stressed the importance of utilizing customer premise equipment such as idle desktops in cafes and handheld devices in proximity, for establishing the fog setup~\cite{chang2017indie}. However, to encourage the participation of such third parties in establishing micro data centres and to make their devices available at the edge, proper incentive mechanisms are necessary. Several incentive models for fog computing are studied in literature. Zeng et al.~\cite{zeng2018incentive}, proposed an incentive model for heterogeneous fog utilizing the framework of contract theory. Luo et al.~\cite{luo2020incentive},  proposed an incentive-aware micro computing cluster formation problem as a coalition game. Similarly, Nazih et al~\cite{nazih2020incentive}, proposed a Stackelberg game-based incentive mechanism for vehicular fog networks. Incentive models were also proposed in similar domains such as in device-to-device (D2D) offloading~\cite{pu2016d2d}. However, most of these incentive models are based on theory and are evaluated only in simulated environments. Development of further real-world adaptable solutions are to be studied, as this is one of the major hurdles in successful deployment of fog computing applications. 

The combination of volatile edge/fog resources and stable cloud resources can reduce the operating costs for some of the cloud services. This opens the potential market to telco operators, who manage the mobile phone infrastructure, through multi-access edge computing. With emerging application areas such as smart city sensing and autonomous vehicles, the telco vendors are likely to form alliances with existing cloud providers for supporting edge analytics and real-time stream data processing~\cite{buyya2018manifesto}. There may be a corollary of this alliance with cloud providers entering the telco domain, which may be perceived as a threat to the main business of the telco providers. 

\subsection{Discussion}

Fog computing and its related challenges are studied extensively in the literature. Reference architectures for establishing fog setups are studied and frameworks were developed to support distributed data processing, ML based edge analytics, serverless data pipelines and streaming data processing on fog infrastructure. The frameworks demonstrated several pilot case studies in different application domains such as smart healthcare, smart cities, interactive games, intelligent transportation systems etc. To study fog computing and its relevant challenges at scale, simulators were developed. Simulators were successfully used for studying and evaluating the resource management policies, scheduling and placement strategies and fog/edge node mobility. Several heuristics, meta-heuristics and ML based solutions were identified for the resource provisioning and fog placement strategies with multiple objectives such as reducing latency, cost, bandwidth usage, and improving energy-efficiency, also considering fog user perspectives such as deadlines and priority of applications. Mobility management is addressed along with solutions for migration and handoff processes. While the results are interesting, adapting the strategies in real-world fog applications is strongly encouraged and the development of relevant fog computing frameworks with proper customer support is essential. 

Regarding QoS provisioning challenges of fog computing, relevant studies dealt with security and privacy, scalability, sustainability, availability and reliability. To address security issues of fog computing, light-weight cryptography-based security protocols with key exchange and homomorphic encryption are developed. Different anomaly-based, ML-based and statistical-based techniques are developed in the literature to address intrusion detection. Blockchain based solutions are developed to address privacy, distributed trust management, and reliability challenges. For future work, to support high mobility devices in fog computing, development of proper handshake and authentication protocols is proposed. In addition, to address energy-efficiency and thus the availability/reliability in fog deployments, future studies should focus on energy harvesters and battery storage. Moreover, the studies should also optimize the location and distance among the fog and edge nodes.

Regarding networking and communication, several long and short ranged, low-powered, low-cost and low-throughput wireless communication technologies are developed. Improvements are also suggested for the wired networks with solutions such as LRPON, to support fog computing. Solutions were also developed in utilizing SDN/NFV for ideally deploying fog services over heterogeneous hardware such as network switches and routers. While there exist several options, choosing ideal communication technologies and protocols for developing/deploying a specific fog application still remains a major challenge.

While fog computing challenges are addressed extensively in the literature, after a decade of research, we still do not see large-scale deployments of public/private fog networks, which can be utilized in realizing interesting IoT applications. Fog economics is not studied to the required extent, while it is the most important challenge for fog adoption. Thus, fog computing did not present a clear business case for the companies and participating individuals yet. Moreover, this type of opportunistic offloading is perceived to be threat to the main business of proximal infrastructure providers such as mobile operators. Furthermore, to utilize proximal fog nodes for opportunistic offloading and to encourage individual/third-party participants, ideal incentive mechanisms are necessary, and literature provides only theoretical incentive models, that too evaluated in simulated environments. Development of further real-world adaptable fog economics solutions are to be studied and tested in large-scale pilot case studies. 

\section{Future research directions for next-generation IoT/fog computing}
\label{sec:futureWork}

Fog computing challenges and the studied solutions are discussed in the previous section. The section already discussed future research directions for each of the considered challenges. In addition to these, we see further scope for research in fog computing in terms of establishing large-scale testbeds and developing dynamic deployment solutions. Moreover, the emerging trends in this domain (e.g. federated learning) and computer science in general (e.g. quantum computing), should positively drive the fog adoption in real-world applications, in the near future. Thus, the following future research directions should help in realizing the objectives of Next Generation Internet of Things (NGIoT) initiative~\cite{NGIoT}, of lowering the barrier for adoption and development of IoT-empowered solutions. 

%Fog computing challenges and the studied solutions are discussed in the previous section. The section already discussed future research directions for each of the considered challenges. In addition to these, we see further scope for research in fog computing through the emerging trends in this domain (e.g. federated learning) and computer science in general (e.g. quantum computing), that should positively drive the fog adoption in real-world applications.

\subsection{Action plan towards large-scale fog computing testbeds and experiments}

Fog computing applications are demonstrated in several small-scale pilot case studies in different domains. One of the main reasons for these small-scale experiments is due to the lack of adequate infrastructure. There are no proper large-scale testbeds, either commercial or academic, on which such prototypes can be demonstrated and evaluated. Such testbeds are tried and common in mobile computing and telecommunication scenarios, with partnership from academic and industrial institutions. For example, Midoglu et al.~\cite{midoglu2021large}, studied the speed of the Mobile Broadband (MBB) networks on Measuring Mobile Broadband Networks in Europe (MONROE) testbed. 

Such efforts are also attempted in cloud computing domain with hybrid/private clouds such as ELIXIR~\cite{crosswell2012elixir} that is used by the bioinformatics and life science services community in Europe. Similar efforts can be planned for establishing fog computing testbeds. With several fog computing applications feasible in IIoT, smart city, autonomous vehicles etc. ecosystems, forming a government, industrial and academic consortium, for establishing large-scale testbeds should be seriously explored. Such efforts may further drive the fog computing research with interoperability, live migration and handover of jobs etc. across the fog networks, which are currently being studied theoretically and demonstrated only on simulators. 

\subsection{Standards-compliant dynamic deployment of fog services}

Currently, fog services are deployed on the resource-constrained and heterogeneous devices, using application-specific proprietary solutions. DevOps is a methodology that integrates and automates software development (Dev) and IT operations (Ops), using relevant tools, thus shortening the systems development life cycle. DevOps is extensively used in the development of cloud services and applications. Standards such as OASIS - Topology and Orchestration Specification for Cloud Applications (TOSCA) are developed that can be used to define the configuration of applications using relevant templates. Utilizing the templates, TOSCA compatible orchestrators can deploy the applications across multiple clouds. This also makes migration of the applications across multiple cloud providers relatively easy~\cite{dehury2022toscadata}. 

Automated deployment and applicability of DevOps for fog computing is not explored that much. TOSCA standard can be extended for supporting fog computing. Container orchestration (e.g. Docker Swarm, Kubernetes) can be used to handle platform independence and interoperability of fog devices. Both the solutions can be combined, and seamless coordination and cooperation can be achieved across fog devices. Early efforts of adapting TOSCA for fog computing are studied in FogDEFT framework~\cite{srirama2022fog}. The framework abstracts all the heterogeneity and complexities and offers a user-friendly paradigm to model and dynamically deploy fog services, on-demand, on the fly, from a remote system. Further work is required to take the approach to the standardization level.

\subsection{Federated learning as a service}

Federated learning (FL) is an emerging ML technique that trains over resource-constrained edge/fog devices using only the local data samples. The locally trained ML model parameters are collected and aggregated at a centralized coordinator and the consolidated models are replaced at the fog nodes for the next iteration of training, until convergence. FL is being used in a wide range of IoT and fog computing case studies such as in smart healthcare, smart cities, autonomous vehicles etc~\cite{pandya2023federated,BANABILAH2022103061}. FL is also addressing several fog computing challenges such as security and privacy~\cite{issa2023blockchain} using solutions like differential privacy (DP) and secure multiparty computation (SMC). Latest developments in this domain include Transfer Learning (training on datasets shared across several participants, which can be fog nodes), Dispersed FL (sub-global model is aggregated within groups which is converged in the second stage, which can realize collaborative learning across multiple layers in the hierarchical fog computing architecture), etc. FL is being extensively explored in recent years.

FL puts additional load on resource-constrained edge/fog devices. Thus, future research in this domain should focus at developing light-weight ML models and optimizing these FL algorithms considering constraints such as local model accuracy, energy requirements, computational resource availability etc~\cite{xiao2021vehicle}. In addition, it is also interesting to study whether FL can be provided as a service from the fog nodes. Does this offer new economic models, thus driving further adoption of fog computing? FL is to be explored further considering all these dimensions.  

\subsection{Quantum cloud computing and its repercussions on edge/fog}

Another rapidly emerging technology is Quantum Computing, which harnesses the laws of quantum mechanics for solving problems too complex for classical computers~\cite{GYONGYOSI201951}. If powerful quantum computers will become widely available in the near future, they may drive the future of quantum cloud computing. These developments will offer solutions such as fault-tolerant secure quantum computations, quantum techniques for cryptographic verification and access control in cloud computing etc ~\cite{gill2022quantum}. Quantum cloud computing clients then need to communicate with the cloud via a quantum link for transferring their tasks and associated qubits. Initial efforts have already been made in this direction~\cite{10064036}. 

With fog computing perceived as cloud in proximity, will the quantum cloud computing have any repercussions on edge/fog computing? Future work should already start exploring the opportunities it is going to offer. For example, integration of Blockchain service with Quantum Internet can improve the communication speed along with the required security provisions for the edge analytics and federated learning tasks performed on the edge-cloud continuum.

\section{CONCLUSION}\label{sec:conclusion}
Fog computing was coined by Cisco, a decade ago, which utilizes proximal computational resources for processing the sensor data, as part of IoT applications. Ever since its proposal, fog computing has attracted significant attention by both research community and industry, and people focused at addressing different challenges/aspects in realizing and utilizing the fog setup. However, after a decade of research, we still do not see large-scale deployments of public/private fog networks, which can be utilized in realizing interesting IoT applications. In the literature, we only see pilot case studies and small-scale testbeds, and utilization of simulators for demonstrating scale of the specified models addressing the respective technical challenges. There are several reasons for this. 

This paper first explored the case studies of fog computing in different domains, justifying the relevance of fog computing. Later it grouped the fog computing challenges such as fog frameworks, simulators, resource management, placement strategies, quality of service aspects, fog economics etc. into different clusters and summarized them along with the state-of-the-art and future research directions. We followed this with a thorough discussion stressing the need for the development of further real-world adaptable fog economics solutions and testing them in large-scale pilot case studies. Later, we outlined the further future research directions in fog computing from the perspective of the emerging trends in this domain and computer science in general, that could lead to the eventual deployment of fog networks and ubiquity of the fog-based applications.

\section*{Acknowledgement}
 This research is supported by SERB, India, through grant CRG/2021/003888. We also thank financial support to UoH-IoE by MHRD, India (F11/9/2019-U3(A)).

%\nocite{*}% Show all bib entries - both cited and uncited; comment this line to view only cited bib entries;
\bibliography{DecadeOfFog}%

\clearpage

% \section*{Author Biography}

% \begin{biography}{\includegraphics[width=66pt,height=86pt,draft]{empty}}{\textbf{Author Name.} This is sample author biography text this is sample author biography text this is sample author biography text this is sample author biography text this is sample author biography text this is sample author biography text this is sample author biography text this is sample author biography text this is sample author biography text this is sample author biography text this is sample author biography text this is sample author biography text this is sample author biography text this is sample author biography text this is sample author biography text this is sample author biography text this is sample author biography text this is sample author biography text this is sample author biography text this is sample author biography text this is sample author biography text.}
% \end{biography}

\end{document}